\documentclass[12pt]{iopart}

\usepackage{iopams}
\usepackage{graphicx}

\newcommand{\cwb}{{\tt coherent WaveBurst}}
\newcommand{\GravEn}{{\tt GravEn}}
\newcommand{\RIDGE}{{\tt RIDGE}}
\newcommand{\xpipeline}{{\tt X-Pipeline}}
\newcommand{\WaveletDetectionFilter}{{\tt Wavelet Detection Filter}}
\newcommand{\WDF}{{\tt WDF}}
\newcommand{\Flare}{{\tt Flare}}

\begin{document}

\title[Astrophysically Triggered Searches for GWs]{Astrophysically Triggered Searches for Gravitational Waves: Status and Prospects}

\author{B.~Abbott$^{16}$,
R.~Abbott$^{16}$, R.~Adhikari$^{16}$, P.~Ajith$^{2}$,
B.~Allen$^{2,54}$, G.~Allen$^{32}$, R.~Amin$^{20}$,
S.~B.~Anderson$^{16}$, W.~G.~Anderson$^{54}$, M.~A.~Arain$^{41}$
% ...
\\
(The LIGO Scientific Collaboration, http://www.ligo.org)}

% ...
\address{$^2$Albert-Einstein-Institut, Max-Planck-Institut f\"ur Gravitationsphysik, D-30167 Hannover, Germany}
% ...
\address{$^{16}$LIGO - California Institute of Technology, Pasadena, CA  91125, USA}
% ...
\address{$^{20}$Louisiana State University, Baton Rouge, LA 70803, USA}
% ...
\address{$^{32}$Stanford University, Stanford, CA 94305, USA}
% ...
\address{$^{41}$University of Florida, Gainesville, FL  32611, USA}
% ...
\address{$^{54}$University of Wisconsin-Milwaukee, Milwaukee, WI 53201, USA}
% ...


\author{F.Acernese $^{7,9}$,
M.Alshourbagy$^{15,16}$, P.Amico$^{13,14}$, F.Antonucci$^{19}$,
S.Aoudia$^{10}$, P.Astone$^{19}$, S.Avino$^{7,8}$, L.Baggio$^1$,
G.Ballardin$^2$, F.Barone$^{7,9}$
% ...
\\
(Virgo Collaboration, http://www.virgo.infn.it/)}

\address{$^1$Laboratoire d'Annecy-le-Vieux de Physique des Particules (LAPP),  IN2P3/CNRS, Universit\'e de Savoie, Annecy-le-Vieux, France}
\address{$^2$European Gravitational Observatory (EGO), Cascina (Pi), Italia.}
% ...
\address{$^7$ INFN, sezione di Napoli, Italia.}
\address{$^8$ Universit\`a di Napoli "Federico II" Complesso Universitario di Monte S.Angelo, Italia.}
\address{$^9$ Universit\`a di Salerno, Fisciano (Sa), Italia.}
\address{$^{10}$Departement Artemis -- Observatoire de la C\^ote d'Azur, BP 4229 06304 Nice, Cedex 4, France.}
% ...
\address{$^{13}$INFN, Sezione di Perugia, Italia.}
\address{$^{14}$Universit\`a di Perugia, Perugia, Italia.}
\address{$^{15}$INFN, Sezione di Pisa, Italia.}
\address{$^{16}$ Universit\`a di Pisa, Pisa, Italia.}
% ...
\address{$^{19}$INFN, Sezione di Roma, Italia.}
\address{$^{20}$Universit\`a "La Sapienza",  Roma, Italia}
% ...

\ead{Z. M\'{a}rka, zsuzsa@astro.columbia.edu}

\begin{abstract}
In gravitational-wave detection, special emphasis is put onto
searches that focus on cosmic events detected by other types of
astrophysical observatories. The astrophysical triggers, e.g. from
$\gamma$-ray and X-ray satellites, optical telescopes and neutrino
observatories, provide a trigger time for analyzing gravitational
wave data coincident with the event. In certain cases the expected
frequency range, source energetics, directional and progenitor
information is also available. Beyond allowing the recognition of
gravitational waveforms with amplitudes closer to the noise floor of
the detector, these triggered searches should also lead to rich
science results even before the onset of Advanced LIGO. In this
paper we provide a broad review of LIGO's astrophysically triggered
searches and the sources they target.
\end{abstract}


\section{Introduction}

Coalescing binaries, supernovae, gamma ray bursts (GRBs), soft gamma
ray repeaters (SGRs) and other transient sources are not only
interesting candidates for gravitational wave (GW) searches but may
also be observed by other means, such as gamma-rays, X-rays, visible
light and neutrinos. Therefore GW searches can take advantage of the
astrophysical events detected by such independent observatories.
Correlation in time (and direction when available) between candidate
events in the LIGO/Virgo detectors~\cite{LIGO,Virgo} and the
astrophysical trigger event can greatly increase the confidence in
the eventual claim of a detection of GWs. Search strategies can be
optimized in this respect~\cite{Arnaud_GW_SN,MohantyMarka2004}.
Furthermore, several long-term goals of GW astrophysics require
detection of astrophysical events in other channels beyond GWs. For
example, any association between short hard GRBs and inspiraling
neutron star binaries may be confirmed in this
manner~\cite{Nakar2007}. The joint detection of GWs and neutrinos
together with the observation of the optical lightcurve from a
nearby supernova would greatly enhance our understanding of
supernova explosions~\cite{2006RPPh...69..971K}. The current
sensitivity of the LIGO and Virgo detectors allows interesting
astrophysical statements to be made by triggered searches for
close-by events (see recent results in~\cite{SGRpaper,070201}). In
this paper we present a brief overview of the strategies employed by
the members of the Externally Triggered Searches group of the LIGO
Scientific Collaboration (LSC) and Virgo Collaboration and some of
the astrophysical sources targeted.

\section{Strategies for Externally Triggered Searches for GWs}

An external trigger provides information about the GW source that
allows us to impose additional requirements on candidate signals in
the GW data.  We can thereby achieve better background rejection and
higher sensitivity to real GW signals.

The first requirement imposed during a triggered search for GWs is
that the candidate signal be coincident in time (within an
astrophysically motivated window) with the external trigger. By
restricting attention to a subset of the available GW data, a
triggered search can be run with a lower event detection threshold
than an un-triggered search, giving a higher detection probability
at a fixed false alarm probability and better limits in the absence
of a detection. Similarly, knowledge of the source direction allows
us to search only a relevant part of the sky or, depending on the
analysis method, veto candidate events seen in multiple detectors at
times not consistent with the expected GW arrival time difference.
In some cases electromagnetic observations contain information on
the expected GW frequency content (see e.g.\,\cite{SGRpaper}), and
thus a frequency-band-specific analysis of the GW data set can be
performed.

External observations indicate specific progenitor source types for
certain trigger events. In such cases model dependent searches for
GWs can be executed. One example is short GRBs, which are thought to
be produced by coalescing compact binary systems, and whose GW
signal can be detected by matched filtering~\cite{070201}. Another
model-dependent search algorithm is based on van Putten's model for
long GRBs~\cite{Raffaipaper,vanPutten}.

\subsection{Methods for externally triggered searches}
\label{methods}

Published LIGO observational results were obtained from
cross-correlation analyses of data from multiple
detectors~\cite{GRB030329,S2S3S4} as well as via a method that uses
data from a single detector and finds excess power in
astrophysically motivated frequency bands~\cite{SGRpaper}.

\textit{Coherent network analysis} methods, currently under
deployment, address the detection and reconstruction of GWs with
networks of detectors
\cite{ArnaudNetwork,Ch_etal:06,KlMoRaMi:05,Ra:06}. Based on aperture
synthesis, they reconstruct the detector responses to maximize the
signal-to-noise ratio of a gravitational wave from a given sky
direction. These reconstructed responses are used to construct
coherent detection statistics which utilize both the excess power
and the cross-correlation energies of the GW signal detected by the
network. By combining data from several GW detectors, the coherent
methods not only take advantage of the known sky location of a
trigger event, but also allow us a consistency test of the events
detected in different detectors. Therefore, the coherent methods are
expected to have better sensitivity at a given false alarm rate than
than approaches that test for coincident triggers from individual
detector searches. Various coherent statistics, such as the null
stream and the network correlation coefficient can be constructed to
distinguish a genuine GW signal from the environmental and
instrumental artifacts.

Several coherent analysis pipelines are now in use. The
\cwb~pipeline is based on the constraint likelihood
method~\cite{KlMoRaMi:05} and performs the coherent network analysis
in the wavelet domain. The
\RIDGE~\cite{ridge,2002CQGra..19.1513M,2003CQGra..20S.925M} pipeline
uses the Tikhonov regularization scheme~\cite{Ra:06}. The
\xpipeline~is a flexible, general-purpose analysis
package~\cite{Ch_etal:06} for coherent network analysis in the
Fourier domain. These pipelines can be used both for the all-sky and
triggered searches for gravitational wave bursts and they provide
complementary evaluations of the data.

To study the efficacy of data analysis algorithms for externally
triggered searches, realistic Monte-Carlo simulations of
astrophysically motivated signals are used.  The
\GravEn~\cite{Graven} simulation engine, which has already been used
in un-triggered searches during the fourth and fifth LIGO science
runs (S4 and S5), is being adapted for use in triggered searches.
{\GravEn} simulates the response to gravitational waves of the three
LIGO detectors as well as GEO, TAMA, and Virgo, and also provides a
variety of diagnostic information for each detector site.

\section{GWs from Gamma Ray Burst Engines} 

Gamma ray bursts (GRBs) are intense flashes of $\gamma$-rays which
are observed to be isotropically distributed over the
sky~\cite{Piran2005,Meszaros2002}. The leading hypothesis for most
short GRBs (lasting less than $\sim$2~s) is the merger of neutron
star or neutron star -- black hole binaries (see~\cite{Nakar2007}
and references therein). Long GRBs are associated with hypernovae
(see e.g.~\cite{hypernova}). In both scenarios the GRB central
engine is an accreting solar-mass black hole, and so it is plausible
that GRB central engines are also strong emitters of GWs. The GW
signal produced by an inspiralling compact binary is well-modeled,
and can therefore be detected by matched filtering
\cite{inspiralS3S4,070201}.  The GW emission from the binary merger
phase and from hypernovae are not well-understood, necessitating the
use of burst-detection techniques for these sources.


\subsection{Online Searches for GWs using GRB triggers}

A near-real time automated analysis was implemented to search LIGO
data for GW bursts around triggers received from the IPN/GCN
network~\cite{IPN,GCN}. This \textit{online} search is based on
cross-correlating data streams from pairs of detectors. Analysis of
LIGO data coincident with 39 GRBs during the second, third and
fourth LIGO science runs (S2, S3, S4) found no associated GW burst
signals \cite{S2S3S4}. According to this published LSC result, for
S4 the best upper limit on the root-sum-square amplitude of a GW
associated with a GRB was $\sim 1\times 10^{-21} \textrm{Hz}^{-1/2}$
for circularly polarized waves at 150 Hz.

During S5, Nov.~2005 -- Oct.~2007, more than 200 GRB triggers were
received. For $\sim$70\% of these GRB triggers, at least 2 LIGO
detectors were operating, and for $\sim$40\%, all three LIGO
detectors were collecting data. This large sample of GRBs is the
basis for an ongoing search for associated GWB signals.

\subsection{Joint LIGO-Virgo Searches for GWs using GRB triggers}

The advent of data sharing between Virgo and LIGO provides the
opportunity to conduct joint searches using high-sensitivity,
non-aligned detectors at three distinct locations. Although the
Virgo detector is currently somewhat less sensitive than the 4~km
LIGO interferometers, Virgo data can already be beneficial for
triggered searches when the Virgo antenna factors are more
advantageous at the received trigger position.

We plan to perform a combined analysis of data in coincidence with
the $\sim$50 GRB triggers received during the joint LIGO-Virgo
data-taking period (May -- Oct.~2007).
This analysis will involve one or more of the coherent network
methods discussed in Section~\ref{methods}.
In addition, the Virgo Collaboration has independently developed a
procedure for GRB triggered searches~\cite{Acernese2007,Corsi2007},
based on the use of the {\WaveletDetectionFilter} (\WDF), a
wavelet-based transient detection tool \cite{Cuoco2005a,Cuoco2005b}.

While the coherent methods are expected on theoretical grounds to be
the most powerful tools for obtaining astrophysically interesting
bounds, the coincidence search using {\WDF} will provide robustness
against possible noise non-stationarities.

\subsection{GRB Population Study} 

While the GW signals from individual GRBs may be too weak to be
detected directly, the small correlations they induce in the data
near the GRB trigger time may still be detectable by statistical
comparison to data from times not associated with a GRB. The
Finn-Mohanty-Romano algorithm (FMR)
\cite{PhysRevD.60.121101,2005CQGra..22S1349M,2006CQGra..23S.723M}
applies a two-sample test on the sets of inter-detector
cross-correlations obtained from times with and without GRB
triggers. The power of the FMR test increases as $N^{1/4}$, where
$N$ is the number of triggers at similar redshift, allowing it to
accumulate signal-to-noise ratio over a population of GRBs. This
algorithm can place upper limits on the population-averaged energy
radiated in GWs \cite{2002ApJ...575..111B}.

We have applied an FMR-inspired algorithm to S2-S3-S4 LIGO
data~\cite{S2S3S4}. For the future we plan to incorporate priors
using the redshift distribution of observed GRBs from astrophysical
literature (see for example
\cite{2002ApJ...575..111B,2005A&A...435..421G}), and apply it to the
S5 GRB set.

\section{Triggers Associated with Soft Gamma-ray Repeaters (SGRs)} 

Soft $\gamma$-ray Repeaters (SGRs) are objects (possibly highly
magnetized neutron stars~\cite{Kouveliotou1999}) that emit
short-duration X- and $\gamma$-ray bursts at irregular intervals.
Occasionally, these objects also emit giant flares lasting hundreds
of seconds with peak electromagnetic luminosities reaching $10^{47}$
erg/s~\cite{Woods2007}. Up to 15\% of short GRBs might be due to
SGRs~\cite{Nakar2006}. Since these flares might be accompanied by
catastrophic non-radial motion in the stellar matter, galactic SGRs
may produce detectable gravitational waves. Furthermore, the X-ray
light curve of some SGR giant flares exhibit quasi-periodic
oscillations (QPOs) (see e.g.~\cite{Israel2005}) at well-defined
frequencies. These QPOs may be due to seismic modes of the neutron
star~\cite{watts1,watts2,watts3,Israel2005} which in turn could emit
GWs.

Search methods have been developed which target both the
instantaneous gravitational emission at the burst,
\Flare~\cite{Kalmuspaper}, and the quasi-periodic seismic
oscillations of the object following the catastrophic cosmic
event~\cite{Luca}.  The QPO analysis has been applied to the
available LIGO data for the December 27, 2004 hyperflare of
SGR1806-20. At the time of the event, the LIGO detectors were under
commissioning in preparation for the S4 science run; only the
Hanford 4~km detector collected data in Astrowatch
mode~\cite{SGRpaper,Astrowatch}~\footnote{Currently the 2 km
interferometer at Hanford (H2), the GEO600 interferometer in Germany
and the Virgo detector in Italy remain operable for periods of time
and participate in the Astrowatch program (A5). H2 can collect data
when its operation does not interfere with Enhanced LIGO
commissioning activities, giving us opportunities to continue
astrophysical observations at ~S5 sensitivity levels at a much
reduced duty factor. GEO600 is also operational, with a high duty
factor, but at a significantly lower sensitivity at low frequencies.
Virgo is undergoing a series of upgrades with long commissioning
periods, during which data of scientific interest could be collected
with much reduced duty factor.}. The best upper limit result by the
LSC, for the 92.5 Hz QPO, corresponds to a GW emission of
7.7~$\times$~10$^{46}$ erg~\cite{SGRpaper}. This is comparable to
the total isotropic energy emitted electromagnetically by the flare,
and close to the theoretically expected maximum emission of
$\sim10^{46}$\,erg
\cite{deFreitasPacheco1998,Ioka2001a,Horvath2005}. According to a
simple isotropic emission model~\cite{SGRpaper}, the minimum
detectable GW energy released by the source scales with the square
of the strain sensitivity. Therefore it is expected, that with over
an order of magnitude of sensitivity increase for advanced GW
detectors, we will be able to probe the energetics for close-by
galactic SGR sources orders of magnitudes below the characteristic
$10^{46}$~erg level.

Several hundred SGR bursts were observed electromagnetically during
S5. Most of these are attributable to SGR1806-20 and SGR1900+14,
both of which have exhibited QPOs in the
past~\cite{watts1,watts2,watts3,Israel2005}. These SGR bursts are
the target of ongoing searches.

\section{Other Sources}

\subsection{Low Mass X-ray Binaries} 

Low mass X-ray binaries (LMXBs) are potential sources of
GWs~\cite{heyl}. In particular, it has been proposed that
\textit{r}-modes inside the neutron star are driven by accretion and
generate GWs \cite{Andersson,Bildsten}. We therefore plan to perform
externally triggered searches for GWs from known LMXBs.
Investigations of the sensitivity of the global detector network to
GW bursts from Sco-X1 are reported in~\cite{LMXB_ScoX}. We also plan
to search for coincidences with X-ray data from the Rossi X-Ray
Timing Explorer (RXTE) satellite~\cite{RXTE}. RXTE can recognize
changes in X-ray brightness that occur on a millisecond timescale,
thus providing crucial trigger information about X-ray bursts in
LMXBs~\footnote{RXTE data can enhance our searches for gravitational
waves by (a.) providing information on transient events, such as
flares from SGRs (b.) providing possible values for parameters of
damped normal modes of the neutron star associated with
quasi-periodic oscillations~\cite{SGRpaper} and (c.) providing X-ray
light curves for Low Mass X-ray Binaries that can be used as
templates in searches for possible gravitational wave signatures in
the interferometric data~\cite{Markowitz}.}. In the absence of a
detection, upper limits can be used to derive constraints on
accretion or the \textit{r}-modes in LXMBs.


\subsection{Pulsar Glitches} 

Pulsar glitches are observed as step increases in the rotation
frequency of pulsars.  The increase in rotational energy is
$\sim$10$^{43}$ erg.  These glitches are likely caused by a
decoupling between the star's solid crust and superfluid interior
(for older pulsars) \cite{Cheng:1988}, or by
reconfigurations of the crust as spin-down reduces the centrifugal
force and the crust reaches breaking strain (for younger pulsars)
\cite{Franco:2000}.  In either case, the disruption should
excite oscillatory modes throughout the neutron star, leading
to the emission of a burst of GWs in the form of a decaying sinusoid, or
`ring-down'~\cite{Thorne:1969}.
The frequencies and decay times of the modes are determined by
the equation of state of the neutron star and its mass and radius;
GW observations may be used to constrain these parameters to
within a few percent \cite{Andersson:1998}.
The time of the sudden spin-up is limited to within $\sim$2 minutes
observationally~\cite{mcculloch}, allowing its use as a search trigger.

A Bayesian search method has been developed~\cite{Clark:2007}, where
the evidence for a ring-down signal is compared with that for a
noise model on data around a pulsar glitch trigger. This method
includes priors for the signal parameters, as inferred from
numerical simulations.  The framework also allows the incorporation
of alternative signal models which can be used to automatically veto
instrumental transients. This method is currently being used to
search for GWs associated with pulsar glitches during S5 (including
the August 12th, 2006 event of PSR B0833-45~\cite{Flanagan:2006}).


\subsection{Neutrinos}

Several astrophysical phenomena, such as core-collapse supernovae,
are expected to emit both GWs and
neutrinos~\cite{2006RPPh...69..971K}. The arrival time of neutrinos
at detectors, such as Super-Kamiokande~\cite{Fukuda03} and
IceCube~\cite{2004APh....20..507A}, can therefore serve as triggers
for LIGO-Virgo searches.  For nearby supernovae,
SNEWS~\cite{Antonioli04} will provide an alert and LIGO/Virgo will
respond by analyzing the data around the reported event time. Apart
from their purely astrophysical interest, GW-neutrino coincidences
from a supernova would provide new information on neutrino
masses~\cite{ArnaudNuMass}.

High energy neutrinos are expected to be emitted along with GWs from
a long GRB if the progenitor is a
hypernova~\cite{1997PhRvL..78.2292W, 1998PhRvL..80.3690V} or a
compact binary merger~\cite{2007NJPh....9...17L}. High energy
neutrinos can provide superior directional information in addition
to event times. Comparing the source direction reconstructed by
neutrino detectors and GW detectors can increase the confidence of a
detection. Such an analysis pipeline is under
development~\cite{aso..GWNu}.

\subsection{Optical Transients and Supernovae}

The current reach of neutrino detectors is limited to our Galactic
neighborhood, thus optical observations are needed to address
extragalactic events. According to theoretical calculations, the
electromagnetic fluxes expected from plausible sources of GWs should
be sufficient to allow the observation of optical counterparts to GW
events~\cite{Sylvestre2003}. Because the light curve of such a
source does not appear immediately, an external trigger derived from
optical observations leads to an uncertainty of several hours in the
trigger time, making the data analysis task more challenging but
still tractable. Since the sky position is well-known, directional
analysis methods are applicable. An interesting alternative is to
use the source location reconstructed from candidate GW events for
follow up observations with optical telescopes in order to seek
confirmation of the event candidate; development of this approach
has started during the summer of 2007~\cite{loocup}.

\bigskip
\noindent There are other astrophysical transients whose connection
to GWs is yet to be explored. For example, blazar flares are powered
by accretion onto a supermassive black hole at the center of the
host galaxy. Similarly to GRBs, they also have a central engine and
jet. Since it has been suggested that some blazars could contain
binary black holes~\cite{BBHblazar}, they may become future objects
of interest for GW searches.

\section{Conclusions}

Interesting results from astrophysically triggered searches using
interferometric GW data have been already
published~\cite{GRB030329,S2S3S4,SGRpaper}. The LIGO detectors have
reached their design sensitivities, which already allow us to make
specific scientific statements for close-by events, e.g.
constraining the source type (or position) of GRB070201, a
short-hard GRB event observed to come from a direction overlapping
M31~\cite{070201}.

With the further improvement of interferometric detectors and the
application of advanced network methods for externally triggered
searches to the LIGO-Virgo network, it is likely that associations
between GWs and their electromagnetic counterparts will be confirmed
during the lifetime of the advanced detectors.


\section{Acknowledgements}

The authors gratefully acknowledge the support of the United States
National Science Foundation for the construction and operation of
the LIGO Laboratory, the Science and Technology Facilities Council
of the United Kingdom, the Max-Planck-Society, and the State of
Niedersachsen/Germany for support of the construction and operation
of the GEO600 detector and the Italian Istituto Nazionale di Fisica
Nucleare and the French Centre National de la Recherche Scientifique
for the construction and operation of the Virgo detector. The
authors also gratefully acknowledge the support of the research by
these agencies and by the Australian Research Council, the Council
of Scientific and Industrial Research of India, the Spanish
Ministerio de Educaci\'on y Ciencia, the Conselleria d'Economia,
Hisenda i Innovaci\'o of the Govern de les Illes Balears, the
Scottish Funding Council, the Scottish Universities Physics
Alliance, The National Aeronautics and Space Administration, the
Carnegie Trust, the Leverhulme Trust, the David and Lucile Packard
Foundation, the Research Corporation, and the Alfred P. Sloan
Foundation. This paper has LIGO Document Number LIGO-P070142-01-Z.

\bigskip



\section{References}

\providecommand{\newblock}{}

\end{document}